\documentclass[aps,prd,eqsecnum,nofootinbib,showpacs,twocolumn]{revtex4-1}

\usepackage{amsmath,amssymb,bm}
\usepackage[dvips]{graphicx}
\usepackage{color}
\usepackage{autobreak}
\allowdisplaybreaks

\newcommand{\be}{\begin{equation}}
\newcommand{\ee}{\end{equation}}
\newcommand{\bea}{\begin{eqnarray}}
\newcommand{\eea}{\end{eqnarray}}

\newcommand{\eq}[1]{Eq.~(\ref{eq:#1})}
\newcommand{\sect}[1]{Sec.~\ref{sec:#1}}
\newcommand{\appen}[1]{Appendix~\ref{sec:#1}}

\newcommand{\del}{\partial}

\newcommand{\calO}{{\cal O}}

\newcommand{\eg}{{\it e.g.}}

\bmdefine{\bmq}{{\bm{q}}}
\bmdefine{\bmk}{{\bm{k}}}
\bmdefine{\bmx}{{\bm{x}}}
\bmdefine{\bmy}{{\bm{y}}}
\bmdefine{\bmr}{{\bm{r}}}
\bmdefine{\bmnabla}{{\bm{\nabla}}}
\bmdefine{\bmA}{ \bm{A} }
\bmdefine{\bmD}{ \bm{D} }

\bmdefine{\bmPhi}{ \bm{\Phi} }
\bmdefine{\bmPsi}{ \bm{\Psi} }
\bmdefine{\bmcalO}{ \bm{\mathcal{O}} }

\newcommand{\calM}{{\cal M}}

\newcommand{\tilx}{\tilde{x}}

\newcommand{\tilphi}{\tilde{\phi}}

\bmdefine{\bmg}{{\bm{g}}}
\bmdefine{\bmR}{{\bm{R}}}

\newcommand{\GR}{G^R}

\newcommand{\lam}{\kappa}

\newcommand{\nza}{n_{z}}
\newcommand{\nzb}{n_{z}'}
\newcommand{\nzc}{n_{z}''}
\newcommand{\npa}{n_{p}}
\newcommand{\npb}{n_{p}'}
\newcommand{\npc}{n_{p}''}

\newcommand{\schrodinger}{Schr\"{o}dinger}

\newcommand{\Fp}{\mathcal{F}_+}
\newcommand{\Fm}{\mathcal{F}_-}
\newcommand{\Fpm}{\mathcal{F}_\pm}

\newcommand{\tilpsi}{\tilde{\psi}}
\newcommand{\tilnu}{\tilde{\nu}}

\newcommand{\bwt}{\begin{widetext}}
\newcommand{\ewt}{\end{widetext}}
\newcommand{\bab}{\begin{autobreak}}
\newcommand{\eab}{\end{autobreak}}

\begin{document}



\title{Nonuniqueness of scattering amplitudes at special points}
\author{Makoto Natsuume}
\email{makoto.natsuume@kek.jp}
\altaffiliation[Also at]{
Department of Particle and Nuclear Physics, 
SOKENDAI (The Graduate University for Advanced Studies), 1-1 Oho, 
Tsukuba, Ibaraki, 305-0801, Japan;
 Department of Physics Engineering, Mie University, 
 Tsu, 514-8507, Japan.}
\affiliation{KEK Theory Center, Institute of Particle and Nuclear Studies, 
High Energy Accelerator Research Organization,
Tsukuba, Ibaraki, 305-0801, Japan}
\author{Takashi Okamura}
\email{tokamura@kwansei.ac.jp}
\affiliation{Department of Physics, Kwansei Gakuin University,
Sanda, Hyogo, 669-1337, Japan}
\date{\today}
\begin{abstract}
We point out little discussed phenomenon in elementary quantum mechanics. In one-dimensional potential scattering problems, the scattering amplitudes are not uniquely determined at special points in parameter space. We examine a few explicit examples. We also discuss the relation with the pole-skipping phenomena recently found in holographic duality. In the holographic pole-skipping, the retarded Green's functions are not uniquely determined at imaginary Matsubara frequencies. It turns out that this universality comes from the fact that the corresponding potential scattering problem has the angular momentum potential.
\end{abstract}
%

\maketitle

\section{Introduction and Summary}

In this paper, we point out little discussed phenomenon in an elementary quantum mechanics. Consider a 1-dimensional potential scattering problems:
\begin{align}
-\del_x^2\psi+V(x)\psi = k^2 \psi~,
%
\end{align}
where we set $\hbar^2/(2m)=1$ for simplicity and $E=:k^2$. We consider 1-dimensional scattering problems with $x>0$. This corresponds to the radial motion problems in 3-dimensions. 

In such a problem, we show that the $S$-matrix is not uniquely determined by appropriately choosing the wave number $k$ (in the complex $k$-plane) and potential parameters. The $S$-matrix is not unique because it  takes the form at the special points:
\begin{align}
S= \frac{0}{0}~.
%
\end{align}
Namely, the residue of a pole vanishes. We call such a phenomenon ``pole-skipping" because the would-be pole is ``skipped." More precisely, near the pole-skipping point, the $S$-matrix typically behaves as
\begin{align}
S \propto \frac{\delta k-i\delta\nu}{\delta k+i\delta\nu}~,
%
\end{align}
where $\nu$ is a parameter of the potential $V$. Thus, near the pole-skipping point, the $S$-matrix is not unique and depends on the slope $\delta k/\delta\nu$ how one approaches the pole-skipping point. We examine a few explicit examples in this paper%
\footnote{The pole-skipping is little discussed in literature as far as we are aware, but it is discussed briefly in Ref.~\cite{newton2}.}.

In particular, we mainly focus on the potentials with angular momentum $\nu:=l+1/2$. In such examples, 
the pole-skipping points are located at 
\begin{align}
\nu = -\frac{n}{2}~, \quad (n=1,2,\cdots)~.
%
\end{align}

The $S$-matrix is not uniquely determined because the wave function is not uniquely determined. In our problems, the point $x=0$ is a regular singularity. So, one can obtain the solution via a power-series expansion around $x=0$:
\begin{align}
\psi = x^\lambda(\psi_0+\psi_1 x+\cdots)~.
%
\end{align}
We show that the power-series expansion takes the form $0/0$ at pole-skipping points.
For example, the $O(x)$ term takes the form $0/0$ at the first pole-skipping point $\nu=-1/2$. 
Similarly, the $O(x^2)$ term becomes $0/0$ at the second pole-skipping point $\nu=-1$. 
In this sense, the wave function is not uniquely determined at pole-skipping points, and this leads to the nonuniqueness of $S$.

We see a peculiar property in an elementary quantum mechanics problem. But there is analogous phenomenon in strongly-coupled quantum field theories recently found using holographic duality or AdS/CFT duality \cite{Maldacena:1997re,Witten:1998qj,Witten:1998zw,Gubser:1998bc} (see, \eg, Refs.~\cite{CasalderreySolana:2011us,Natsuume:2014sfa,Ammon:2015wua,Zaanen:2015oix,Hartnoll:2016apf}). Holographic duality is a powerful tool to compute strongly-coupled systems%
\footnote{Readers who are interested only in quantum mechanics problems may skip the following paragraphs and \sect{holography}.}. Quantum field theory is hard to solve at strong coupling. For example, one would like to compute the retarded Green's functions, but it is difficult to compute them at strong coupling. However, holographic duality enables one to obtain them by solving classical gravitational problems. 

Recently, using holographic duality, it is shown that finite-temperature retarded Green's functions are not uniquely determined at special points in the complex momentum space $(\omega,q)$, where $\omega$ is frequency and $q$ is wave number \cite{Grozdanov:2019uhi,Blake:2019otz,Natsuume:2019xcy}. Such a phenomenon is collectively known as pole-skipping. 
Just like the quantum mechanics problem, the Green's function takes the form $G^R=0/0$. More precisely, the Green's function is not uniquely determined. Near the pole-skipping point, the Green's function typically behaves as
\begin{align}
\GR \propto \frac{\delta \omega+\delta q}{\delta \omega-\delta q}~,
\label{eq:pole_skip_example}
\end{align}
So, it depends on the slope $\delta q/\delta\omega$ how one approaches the pole-skipping point. 

The holographic pole-skipping shows a universal behavior. 
The pole-skipping points are always located at Matsubara frequencies. Typically, they are located at%
\footnote{More precisely, the first pole-skipping point
is related to the spin-$s$ of boundary operators as $\omega=(s-1)(2\pi T)i $.}
\begin{align}
\omega=-(2\pi T)ni~, \quad(n=1,2,\cdots)~,
%
\end{align}
where $T$ is temperature.
On the other hand, the value of $q$ depends on the system, but it is complex in general. 

The pole-skipping in holographic duality and the pole-skipping in quantum mechanics are not just mere analogy. 
We show that the holographic pole-skipping problem can be written as a quantum mechanical problem with angular momentum. The universality of the pole-skipping points $\omega$ is translated into the universality of the pole-skipping points in $\nu$.

The plan of the present paper is as follows. 
In \sect{power-series}, we utilize the power-series expansion to identify possible locations where the wave function is not uniquely determined. In \sect{examples}, we examine a few explicit examples where analytic solutions are available and locate the pole-skipping points where the $S$-matrix is not uniquely determined. The results are consistent with the power-series expansion method. In \sect{holography}, we discuss the relation with the holographic pole-skipping. 

\section{Power-series expansion}\label{sec:power-series}
\subsection{$x=0$ expansion}

We consider analytically solvable examples in \sect{examples}, but it is worthwhile to consider a generic potential scattering problem. The discussion below is also useful for the potentials where analytic solutions are not available. We consider
\begin{subequations}
\label{eq:eom_x=0}
\begin{align}
0 &=-\del_x^2\psi+V\psi-k^2\psi~,\\
V &= \frac{\nu^2-1/4}{x^2}+\sum_{n=-1}v_n x^n~,
%
\end{align}
\end{subequations}
where we included the ``angular momentum" part $\nu:=l+1/2$.

The point $x=0$ is a regular singularity, so one can solve the problem by a power-series expansion:
\begin{align}
\psi(x)=\sum_{n=0} \psi_n x^{n+\lambda}~.
%
\end{align}
Near $x\to0$, the angular momentum part dominates:
\begin{align}
0 \sim -\del_x^2\psi + \frac{\nu^2-1/4}{x^2} \psi~,
%
\end{align}
so the solution is
\begin{align}
\psi \sim x^{\lambda_\pm}~, \quad \lambda_\pm := \frac{1}{2}\pm\nu~.
%
\end{align}
For physical angular momentum, $\nu\geq1/2$, so we choose $\lambda_+$. 
More generally, one would impose the boundary condition that the wave-function is square-integrable or $\lambda_+$ with $\nu>-1$.

The coefficient $\psi_n$ is obtained by a recursion relation. At the lowest order, 
\begin{align}
0=M_{11}\psi_0-(1+2\nu)\psi_1~.
%
\end{align}
Normally, this equation determines $\psi_1$ from $\psi_0$. However, when $\nu=-1/2$ and $M_{11}=0$, both $\psi_0$ and $\psi_1$ are free parameters. In this case, the wave function is not uniquely determined but depends on $\psi_1/\psi_0$. $M_{11}$ typically depends on $k^2$ and $v_n$, so the solution $M_{11}=0$ depends on these parameters. Thus, the nonuniqueness occurs at special points of these parameters and $\nu=-1/2$. This determines a pole-skipping point.
As the result of the nonuniqueness of the wave function, the $S$-matrix is not unique. We see explicit examples in next section. 

In the power-series expansion, one generates two independent solutions either by choosing $\lambda_+$ or by choosing $\lambda_-$. However, when two roots $\lambda_+$ and $\lambda_-$ differ by an integer or $\lambda_+-\lambda_-=2\nu$ is an integer, the smaller root fails to produce the independent solution, and the recursion relation breaks down at some order. What we see above is this situation. (In this case, $\nu=-1/2$, so $\lambda_+$ is the smaller root.) 


The nonuniqueness of the wave function continues at higher orders in the recursion relation.  One can write the recursion relation in a matrix form:
\begin{subequations}
\begin{align}
0 & = M\psi \\
& =\begin{pmatrix} 
    M_{11} & -(1+2\nu) & 0 &\cdots \\
    M_{21} & M_{22} & -2(2+2\nu) & \cdots \\
    M_{31} & M_{32} & M_{33} & \cdots \\
    \cdots & \cdots & \cdots & \cdots 
  \end{pmatrix}
  \begin{pmatrix} 
    \psi_0 \\ \psi_1 \\ \psi_2 \\ \cdots
  \end{pmatrix}~.
\label{eq:recursion_matrix}
\end{align}
\end{subequations}
The matrix $\calM^{(n)}$ is obtained by keeping the first $n$ rows and $n$ columns of $M$. The wave function is not uniquely determined when 
\begin{subequations}
\begin{align}
\nu_n &= - \frac{n}{2}~, \quad (n=1,2,\cdots) \\
\det \calM^{(n)}(\nu_n) &= 0~.
\label{eq:condition}
\end{align}
\end{subequations}
Alternatively, one can solve the recursion relation iteratively and analyze the pole-skipping points.

\subsection{$x=\infty$ expansion}

In the examples below, we consider exponentially decaying potentials as $x\to\infty$:
\begin{align}
V \sim e^{-x} +\cdots~.
%
\end{align}
Then, the \schrodinger\ equation typically has a regular singularity at $x=\infty$ (by appropriately changing variables), so let us consider a power-series expansion around $x=\infty$. In this case, $\psi$ asymptotically behaves as
\begin{align}
\psi \sim e^{\pm ikx}~.
%
\end{align}

In order to analyze the point $x=\infty$, set $\tilx=e^{-x}$ and consider the $\tilx=0$ behavior. By redefining
\begin{align}
\psi =: \tilx^{-1/2}\tilpsi~,
%
\end{align}
the \schrodinger\ equation is transformed as 
\begin{subequations}
\begin{align}
0 &= -\del_{\tilx}^2\tilpsi+ \left(\frac{\tilnu^2-1/4}{\tilx^2}+\tilde{V} \right) \tilpsi~,\\
\tilde{V} &= \frac{V}{\tilx^2} = \frac{1}{\tilx}+\cdots~,\\
\tilnu^2 &:= -k^2~.
%
\end{align}
\end{subequations}
Thus, the problem reduces to the power-series expansion problem near $x=0$, where $\nu$ is replaced by $\pm ik$. The boundary condition also reduces to
\begin{align}
\tilpsi \sim \tilx^{1/2+\tilnu}~.
%
\end{align}

Following the same argument as the $x=0$ expansion, one can write a recursion relation. If one chooses $\tilnu=-ik$,
\begin{subequations}
\begin{align}
0 & = M\psi \\
& =\begin{pmatrix} 
    M_{11} & -(1-2ik) & 0 &\cdots \\
    M_{21} & M_{22} & -2(2-2ik) & \cdots \\
    M_{31} & M_{32} & M_{33} & \cdots \\
    \cdots & \cdots & \cdots & \cdots 
  \end{pmatrix}
  \begin{pmatrix} 
    \psi_0 \\ \psi_1 \\ \psi_2 \\ \cdots
  \end{pmatrix}~.
%
\end{align}
\end{subequations}
Thus, in this case, the nonuniqueness occurs at discrete values of $k$:
\begin{align}
ik_n=\frac{n}{2}~, \quad (n=1,2,\cdots)~.
%
\end{align}
Similarly, if one chooses $\tilnu=ik$, the nonuniqueness occurs at
\begin{align}
ik_n=-\frac{n}{2}~.
%
\end{align}

\subsection{Summary}

We identify possible locations where the wave function is not uniquely determined:
\begin{enumerate}
\item
If a potential has the ``angular momentum" part and has a regular singularity at $x=0$, then the wave function is not unique at $\nu=-n/2$ with appropriate $k$ and parameters of $V$. 
\item 
If a potential exponentially decays asymptotically and has a regular singularity at $x=\infty$, then the wave function is not unique at $ik=\pm n/2$  with appropriate parameters of $V$.
\end{enumerate}
We confirm this observation using explicit examples where analytic solutions are available. As the result of the nonuniqueness, we will see that the $S$-matrix is not uniquely determined at these points.

\section{Examples}\label{sec:examples}

If $V$ vanishes fast enough asymptotically $x\to\infty$, the asymptotic solution of the \schrodinger\ equation is
\begin{align}
\psi \sim \Fp e^{-ikx} - \Fm e^{ikx}~,\quad (x\to\infty)~.
\label{eq:asymptotic}
\end{align}
The first term represents the incoming wave, and the second term represents the outgoing wave. The normalization of $\psi$ is specified at $x=0$ [see \eq{bc_x=0_text} and \appen{preliminaries}]. The functions $\mathcal{F}_\pm$ are called the ``Jost functions." Then, the $S$-matrix is given by
\begin{align}
S = e^{2i\delta} = \frac{\Fm}{\Fp}~.
%
\end{align}

We consider three examples below:
\begin{enumerate}
\item A potential that has a regular singularity at $x=0$, 
\item A potential that has regular singularities at $x=0$ and $x=\infty$, 
\item A potential that is regular at $x=0$ but has a regular singularity at $x=\infty$.
\end{enumerate}

\subsection{One-pole approximation}

As a warm-up exercise, consider a simple $S$-matrix problem. The $S$-matrix satisfies the following conditions:
\begin{enumerate}
\item The existence of a pole.
\item $|S|=1$ for real $k$.
\item $S=1$ for $k=0$.
\end{enumerate}
Suppose that $k$ is close to a pole and the other poles are far away. Then, the simplest function which satisfies these conditions is
\begin{align}
S = - \frac{k+ic}{k-ic}~,
%
\end{align}
where the parameter $c$ is determined by an explicit form of the potential. 

For physical scattering, $k$ is real, but make $k$ complex. The $S$-matrix has a pole at
\begin{align}
k=ic~.
%
\end{align} 
If $c>0$, the pole lies on the positive imaginary axis. Such a pole represents a bound state. To see this, write $k=i\kappa (\kappa>0)$. Then, the asymptotic behavior becomes
\begin{align}
\psi \sim \Fp e^{\kappa x} - \Fm e^{-\kappa x}~, \quad (x\to\infty)~.
%
\end{align}
So, the wave function is normalizable and is localized if $\Fp=0$. Namely, a pole of $S$ located on the positive imaginary axis in the complex $k$-plane corresponds to a bound state. 

On the other hand, if $c<0$, the pole is located on the negative imaginary axis. Such a pole is called an ``antibound state" or a ``virtual state." The wave function of an antibound state is not normalizable and diverges at $x\to\infty$. Thus, the physical interpretation is rather obscure, but it has a physical effect. An antibound state causes a large cross section at low energy \cite{bohm}.

A bound state and an antibound state correspond to $\Fp=0$. In addition, the $S$-matrix can diverge when $\Fm=\infty$. These poles are called ``redundant poles" or ``false poles" \cite{redundant}. They do not correspond to physical states. From the symmetry of the $S$-matrix $S(k)=1/S(-k)$, there are corresponding redundant zeros. When one identifies physical poles, it is useful to use the Jost functions $\mathcal{F}_\pm$ instead of the $S$-matrix itself.

Now, $k=c=0$ is the pole-skipping point in the sense that $0/0$ appears. In particular, the pole-skipping means that the $k\to0$ limit and $c\to0$ limit do not commute: 
\begin{itemize}
\item If one takes $c\to0$ limit first, $S=-1$. 
\item If one takes $k\to0$ limit first, then $S=1$.
\end{itemize}
One can understand this from Levinson's theorem: the phase shift at zero energy $\delta(0)$ is related to the number of bound states. For a single bound state, $\delta(0)=\pi$.
So, this pole-skipping is well-understood. However, what we would like to show below is that a similar phenomenon can occur at nontrivial parameters.

\subsection{Coulomb potential}

We first consider the Coulomb potential with  the ``angular momentum" $\nu:=l+1/2$: 
\begin{align}
0=-\del_x^2\psi+ \left( \frac{\nu^2-1/4}{x^2}+\frac{e^2}{x}-k^2 \right)\psi~.
%
\end{align}
The equation has a regular singularity at $x=0$ and an irregular singularity at $x=\infty$.
The equation takes the form of \eq{eom_x=0}.
%
%
%
%
We denote the solution which satisfies the $x=0$ boundary condition as $\varphi$:
\begin{align}
\varphi \sim x^{1/2+\nu}~, \quad (x\to0)~.
\label{eq:bc_x=0_text}
\end{align}

The Coulomb potential is a long-range force which changes the asymptotic behavior \eqref{eq:asymptotic}. Set the ansatz
\begin{align}
\psi \sim e^{\pm ikx+g(x)}~.
%
\end{align}
The \schrodinger\ equation behaves as
\begin{subequations}
\begin{align}
\pm 2ik g' &\sim \frac{e^2}{x} \\
\to g &\sim \mp \frac{ie^2}{2k}\ln x \\
\to \psi &\sim e^{\pm i(kx-\kappa\ln x)}
%
\end{align}
\end{subequations}
where $\kappa:=e^2/(2k)$.
Thus, we choose the Jost functions as 
\begin{align}
\psi \sim \Fp e^{-ikx+i\kappa\ln x} -  \Fm e^{ikx-i\kappa\ln x}~, 
%
\end{align}
as $x\to\infty$. 

Changing the variable
\begin{align}
\zeta=2ikx~,
%
\end{align}
the \schrodinger\ equation reduces to the Whittaker differential equation: 
\begin{align}
0=\del_\zeta^2\psi + \left( -\frac{1}{4} + \frac{i\kappa}{\zeta} - \frac{\nu^2-1/4}{\zeta^2} \right)~.
%
\end{align}
The Whittaker functions
\begin{align}
W_{i\kappa,\nu}(\zeta)~, W_{-i\kappa,\nu}(-\zeta)~, M_{i\kappa,\nu}(\zeta)~, M_{i\kappa,-\nu}(\zeta)~, 
%
\end{align}
are the solutions of the Whittaker differential equation. The function $M_{i\kappa,\nu}(\zeta)$ is ill-defined when $2\nu$ is a negative integer. When $2\nu\neq\mathbb{Z}$, $W_{i\kappa,\nu}(\zeta)$ and $W_{-i\kappa,\nu}(-\zeta)$ [or $M_{i\kappa,\nu}(\zeta)$ and $M_{i\kappa,-\nu}(\zeta)$] are independent solutions. We choose $M_{i\kappa,\nu}(\zeta)$ and $W_{i\kappa,\nu}(\zeta)$ as independent solutions.
The function $M_{i\kappa,\nu}$ behaves as
\begin{align}
M_{i\kappa,\nu} \sim \zeta^{1/2+\nu}~, \quad (\zeta\to0)~.
%
\end{align}
So, the solution which satisfies the $x=0$ boundary condition is
\begin{align}
\varphi = \frac{1}{(2ik)^{\nu+1/2}} M_{i\kappa,\nu}~.
%
\end{align}
As $\zeta\to\infty$, 
\bwt
\begin{subequations}
\begin{align}
\varphi &\sim \frac{1}{(2ik)^{\nu+1/2}} \Gamma(2\nu+1) \left\{ 
\frac{ e^{i\pi(\nu+1/2)}(2e^{-i\pi/2}k)^{i\kappa} }{ \Gamma(\nu+\frac{1}{2}+i\kappa) } e^{-ikx+i\kappa\ln x}
+ \frac{ (2e^{i\pi/2}k)^{-i\kappa} }{ \Gamma(\nu+\frac{1}{2}-i\kappa) } e^{ikx-i\kappa\ln x}
\right\} \\
&=: \frac{1}{-2ik} \{ \Fp e^{-ikx+i\kappa\ln x} - \Fm e^{ikx-i\kappa\ln x} \}~,
%
\end{align}
\end{subequations}
\ewt
when $-\pi/2<\arg\zeta<3\pi/2$.

The Jost function is given by
\begin{align}
\Fp = (2e^{-i\pi/2}k)^{1/2-\nu+i\kappa} \frac{ \Gamma(2\nu+1) }{ \Gamma(\nu+\frac{1}{2}+i\kappa) }~.
%
\end{align}
The $S$-matrix is given by
\begin{align}
S = e^{-i\pi(\nu-1/2)}(2k)^{-2i\kappa} \frac{ \Gamma(\nu+\frac{1}{2}+i\kappa) }{ \Gamma(\nu+\frac{1}{2}-i\kappa) }~.
%
\end{align}
The $S$-matrix has a branch point at $k=0$. 

Let us analyze the pole-skipping for the Coulomb problem. Just like the one-pole approximation example, we study the behavior of $S$ as we vary parameters. At pole-skipping points, a pole and a zero occur simultaneously, and one gets $0/0$. So, we first locate poles and zeros of $S$. 

The $S$-matrix has poles at
\begin{align}
\nu+\frac{1}{2}+i\kappa = -\npa~, 
\label{eq:coulomb_pole}
\end{align}
($\npa =0, 1, \cdots$) or
\begin{align}
k= -\frac{ie^2}{2\npa+2\nu+1}~.
%
\end{align}
This is the Regge trajectory. Whether the pole corresponds to a bound state or an antibound state depends on the sign of $e^2$. If $e^2<0$ or if the potential is attractive, one has a bound state (for $2\npa+2\nu+1>0$) . The corresponding energy eigenvalue is
\begin{align}
E=k^2= -\frac{e^4}{4n^2}~, \quad n=\npa+\nu+\frac{1}{2}~.
%
\end{align}
This is the familiar hydrogen spectrum. On the other hand, if $e^2>0$ or if the potential is repulsive, one has an antibound state.

The $S$-matrix has zeros at
\begin{align}
\nu+\frac{1}{2}-i\kappa = -\nza~, 
\label{eq:coulomb_zero}
\end{align}
%
%
($\nza =0, 1, \cdots$). The pole-skipping occurs at
\begin{subequations}
\begin{align}
\nu &= -\frac{\npa+\nza+1}{2}~, \\
\kappa &= \frac{\nza-\npa}{2i}~.
\label{eq:coulomb_pole_skip}
\end{align}
\end{subequations}
Thus, the pole-skipping occurs but it occurs in unphysical region of the angular momentum $\nu=l+1/2<0$. 
The first few pole-skipping points are
\begin{subequations}
\begin{align}
\nu =-\frac{1}{2}~, \quad
& \kappa=0~, \\
\nu =-1~, \quad
& \kappa =\pm\frac{i}{2}~, \\
\nu =-\frac{3}{2}~, \quad
& \kappa = \pm i~, 0~, \\
\cdots \quad & \cdots 
\nonumber
%
\end{align}
\end{subequations}
Note that the pole-skipping occurs at discrete values of $\nu$, and $\nu$ is evenly spaced. We see the holographic interpretation in \sect{holography}. In holography, the pole-skipping occurs at $\omega=-(2\pi T)ni$, so $\omega$ is evenly spaced as well. We show that $\nu$ corresponds to $-i\omega/(4\pi T)$ in the holographic context. 

The $S$-matrix has the slope-dependence near pole-skipping points. For example, near $(\nu,\kappa)=(-1,i/2)$,
\begin{align}
S \sim (-2ik) \frac{ \delta\nu-i\delta\kappa }{ \delta\nu+i\delta\kappa }~.
\label{eq:coulomb_slope}
\end{align}

The $S$-matrix is not uniquely determined at pole-skipping points because the wave function is not uniquely determined. This can be seen by a power-series expansion in \sect{power-series}, but an analytic solution is available, so it is enough to expand the solution. In this case, the pole-skipping occurs at discrete values of $\nu$, so the pole-skipping can be seen in the $x=0$ expansion. The solution is expanded as
\begin{align}
\varphi = x^{\nu+1/2} \left\{1 + \frac{2k\kappa}{1+2\nu}x 
- \frac{k^2(1+2\nu-4\kappa^2)}{4(1+2\nu)(1+\nu)} x^2 + \cdots \right\}~.
%
\end{align}
The power-series expansion indeed takes the form $0/0$ at pole-skipping points%
\footnote{Strictly speaking, this shows that the Whittaker function is ill-defined at pole-skipping points, but this is not really a problem because our real interest is the $S$-matrix such as \eq{coulomb_slope} which is slightly away from the pole-skipping points.}. 
For example, the $O(x)$ term and higher order terms take the form $0/0$ at the first pole-skipping point $\nu=-1/2$. 
Similarly, the $O(x^2)$ term and higher order terms become $0/0$ at the second pole-skipping points $\nu=-1$. 
In general,  the $O(x^n)$ term is not unique at pole-skipping points $\nu=-n/2$, and there are $n$ pole-skipping points. 

\subsection{P\"{o}schl-Teller potential 1}

As the next example, we consider a P\"{o}schl-Teller potential \cite{Poschl:1933zz}:
\begin{align}
0=-\del_x^2\psi+ \left( \frac{\nu^2-1/4}{\sinh^2 x} -k^2 \right)\psi~.
%
\end{align}
There is actually a large class of exponential type potentials which includes the above potential as a special case \cite{Boonserm:2010px}. The class includes the Eckart potential, the Morse potential, and so on, and it is widely discussed in literature. The $1/\cosh^2x$ potential is discussed in textbooks (see, \eg, \cite{LL}) and is discussed in next subsection. 

The equation has regular singularities at $x=0$ and $x=\infty$ (by appropriately changing variables).
The potential has the same behavior as the Coulomb potential near $x\to0$, so we impose the boundary condition
\begin{align}
\varphi \sim x^{1/2+\nu}~, \quad (x\to0)~.
\label{eq:bc_PT}
\end{align}

Introduce a new variable 
\newpage
\begin{align}
u:=\tanh x
%
\end{align}
($x: 0\to\infty, u:0\to1$). 
As $x\to\infty$, $u\sim 1-2e^{-2x}$, so the incoming wave and the outgoing wave behave as
\begin{subequations}
\begin{align}
\psi_\text{in} &\sim e^{-ikx} \sim \left( \frac{1-u}{2} \right)^{ik/2}~, \\
\psi_\text{out} &\sim e^{ikx} \sim \left( \frac{1-u}{2} \right)^{-ik/2}~.
%
\end{align}
\end{subequations}
The solution which satisfies the $x\to0$ boundary condition is given by a hypergeometric function:
\begin{subequations}
\begin{align}
\varphi &= (1-u^2)^{-ik/2}u^{\nu+1/2}{}_2F_1\left(a,b,c; u^2\right)~,\\
a &= \frac{1}{4}-\frac{1}{2}ik+\frac{1}{2}\nu~,\\
b &=\frac{3}{4}-\frac{1}{2}ik+\frac{1}{2}\nu~,\\
c &= 1+\nu~.
%
\end{align}
\end{subequations}
The hypergeometric function is ill-defined when $c=0,-1,\cdots$.

In order to extract the asymptotic behavior $u\to1$, it is convenient to use the following formula:
\bwt
\begin{align}
{}_2F_1(a,b,c;u^2) = & 
\frac{ \Gamma(c)\Gamma(a+b-c) }{ \Gamma(a)\Gamma(b)} 
(1-u^2)^{c-a-b} {}_2F_1(c-a, c-b, 1+c-a-b ;1-u^2)
\nonumber \\
&+\frac{ \Gamma(c)\Gamma(c-a-b) }{ \Gamma(c-a)\Gamma(c-b)} 
{}_2F_1(a,b,1+a+b-c ;1-u^2)~.
\label{eq:Gauss_transf}
%
\end{align}
\ewt
Then, as $u\to1$,
\begin{subequations}
\begin{align}
\varphi \sim& \frac{2^{\nu+1/2}}{2\sqrt{\pi}} \Gamma(1+\nu)
\biggl\{ 
\frac{\Gamma(-ik)}{ \Gamma(\nu+\frac{1}{2}-ik) } e^{-ikx} 
\nonumber \\
&\hspace{2.5truecm}
+ \frac{\Gamma(ik)}{ \Gamma(\nu+\frac{1}{2}+ik) } e^{ikx} 
\biggr\}
\\
=:& \frac{1}{-2ik} \{ \Fp e^{-ikx} - \Fm e^{ikx} \}~,
%
\end{align}
\end{subequations}
where we use 
\begin{align}
\Gamma(2z) = \frac{2^{2z}}{2\sqrt{\pi}}\Gamma(z)\Gamma(z+\frac{1}{2})~.
%
\end{align}

The Jost function is given by
\begin{align}
\Fp = \frac{2^{\nu+1/2}}{\sqrt{\pi}} \frac{\Gamma(1+\nu)\Gamma(1-ik)}{ \Gamma(\nu+\frac{1}{2}-ik) }~.
%
\end{align}
The $S$-matrix is given by
\begin{align}
S &= - \frac{ \Gamma(\nu+\frac{1}{2}-ik) }{ \Gamma(\nu+\frac{1}{2}+ik) }\frac{\Gamma(ik)}{\Gamma(-ik)}~.
%
\end{align}
We first locate poles and zeros of $S$.

The $S$-matrix has poles at
\begin{subequations}
\begin{alignat}{2}
\text{Pole 1: } &\Gamma(\nu+\frac{1}{2}-ik) &&\to k = -i \left(\npa+\nu+\frac{1}{2}\right)~,  
\\
\text{Pole 2: } &\Gamma(ik) &&\to k = i(\npb+1)~,
%
\end{alignat}
\end{subequations}
($\npa,\npb =0, 1, \cdots$). When $\nu>0$, the former gives an infinite number of antibound states. The latter gives an infinite number of redundant poles in the upper-half plane. 

The $S$-matrix has zeros at
\begin{subequations}
\begin{alignat}{3}
\text{Zero 1: } &\Gamma(\nu+\frac{1}{2}+ik) &&\to k = i \left(\nza+\nu+\frac{1}{2}\right)~,
 \\
\text{Zero 2: } &\Gamma(-ik) &&\to k = -i(\nzb+1)~,
%
\end{alignat}
\end{subequations}
($\nza,\nzb =0, 1, \cdots$).
The latter gives an infinite number of redundant zeros in the lower-half plane.

Because there are two sets of poles and zeros, the pole-skipping analysis is more involved (\appen{details}). 
The $S$-matrix satisfies $S(k)=1/S(-k)$. Thus, if the pole-skipping occurs at $k=k_*$, the pole-skipping also occurs at $k=-k_*$. The pole-skipping occurs in pairs at $k=\pm k_*$. So, it is enough to analyze the pole-skipping in the upper-half $k$-plane. Zero 2 is located in the lower-half $k$-plane, so we do not consider it. 
The pole-skipping points are given as follows:
\begin{enumerate}
\item[(i)] Pole 1 and Zero 1:
\begin{subequations}
\begin{alignat}{3}
\nu &= -n-1~, 
\\
k &= i\left(n-\npa+\frac{1}{2}\right)~, 
%
\end{alignat}
\end{subequations}
($n=0,1,\cdots$ and $\npa<n+1/2$). For example,
\begin{subequations}
\begin{align}
\nu = -1~, \quad
& 2k = i, \\
\nu = -2~, \quad
& 2k = i, 3i, \\
\cdots \quad & \cdots 
\nonumber
%
\end{align}
\end{subequations}
Like the Coulomb potential example, the pole-skipping occurs at discrete values of $\nu$, and $\nu$ is evenly spaced. 
Unlike the Coulomb potential, $\nu$ here is not the angular momentum, but the $\nu<0$ case is singular under the boundary condition \eqref{eq:bc_PT}. So, the pole-skipping does not occur in physical region.

In addition,
\begin{align}
\nu=-\npa-\frac{1}{2}~, \quad k=0~.
%
\end{align}
The pole-skipping again occurs when $\nu<0$, but $\nu=-1/2$ or $\npa=0$ is square-integrable under the boundary condition \eqref{eq:bc_PT}.

%
%
%
%

\item[(ii)] Pole 2 and Zero 1:
\begin{subequations}
\begin{align}
k &=i(\npb+1)~,
\\
\nu &= n-\frac{1}{2}~, 
%
\end{align}
\end{subequations}
($-\npb \leq n \leq \npb+1$). For example,
\begin{subequations}
\begin{align}
k= i~, \quad
&\nu=\pm \frac{1}{2}~, \\
k= 2i~, \quad
& \nu = \pm\frac{1}{2}~,\pm\frac{3}{2}~, \\
\cdots \quad & \cdots 
\nonumber
%
\end{align}
\end{subequations}
The pole-skipping occurs at discrete values of imaginary $k$, and $k$ is evenly spaced. 
In this case, the pole-skipping occurs when $\nu>0$, but Pole 2 is a redundant pole, and it does not corresponds to a physical state, so the physical interpretation may be subtle (see \sect{discussion} and \appen{cutoff}). 

\end{enumerate}

Again, the $S$-matrix has the slope-dependence near pole-skipping points. For example, near a Pole~1-Zero~1 pole-skipping point $(\nu,k)=(-1,i/2)$,
\begin{align}
S = 2\frac{\delta k - i\delta \nu}{\delta k+i\delta \nu}~.
%
\end{align}
Near a Pole~2-Zero~1 pole-skipping point $(k,\nu)=(i,1/2)$,
\begin{align}
S = \frac{\delta k - i\delta \nu}{\delta k}~.
%
\end{align}

The $S$-matrix is not uniquely determined because the wave function is not uniquely determined. This can be seen by a power-series expansion in \sect{power-series}. For Pole 1-Zero 1, the pole-skipping occurs at discrete values of $\nu$, so one expects that this pole-skipping can be seen in the $x=0$ expansion. Near $x\to0$, the solution is expanded as 
\begin{align}
\varphi \sim x^{\nu+1/2} \left\{ 1 + \frac{1-12k^2-4\nu^2}{48(\nu+1)} x^2 +\cdots \right\}~.
%
\end{align}
The power-series expansion indeed takes the form $0/0$ at pole-skipping points for Pole 1-Zero 1. For example, $O(x^2)$ term takes the form $0/0$ at the first pole-skipping points $(\nu,k)=(-1,\pm i/2)$.

For Pole 2-Zero 1, the pole-skipping occurs at discrete values of $k$, so it is appropriate to look at the $x=\infty$ expansion. It is convenient to use a solution which satisfies the $x=\infty$ boundary condition, \eg,  $e^{-ikx}$. Such a solution $f_-$ is given by
\begin{subequations}
\begin{align}
f_- =& (1-e^{-2x})^{1/2-\nu} e^{-ikx} 
\nonumber \\
& \times {}_2F_1(1/2-\nu,1/2-\nu+ik, 1+ik; e^{-2x}) 
\\
\sim& e^{-ikx} \left\{ 1- \frac{i(4\nu^2-1)}{4(k-i)}e^{-2x}+\cdots \right\}~.
%
\end{align}
\end{subequations}
The power-series expansion around $x=\infty$ indeed takes the form $0/0$. The $O(e^{-2x})$ term takes the form $0/0$ at the first pole-skipping point $(k,\nu)=(i,\pm1/2)$.

\subsection{P\"{o}schl-Teller potential 2}

Finally, we consider another type of P\"{o}schl-Teller potential:
\begin{align}
0=-\del_x^2\psi+ \left\{ -\frac{\lam(\lam-1)}{\cosh^2 x} -k^2 \right\}\psi~.
%
\end{align}
The equation is regular at $x=0$, and it corresponds to $\nu=1/2$. But it has a regular singularity at $x=\infty$ (by appropriately changing variables), so it is appropriate to look at the $x=\infty$ expansion and to use the solutions which satisfy the $x\to\infty$ boundary conditions $e^{\pm ikx}$.

Introduce a new variable 
\begin{align}
u:=(1-\tanh x)/2
%
\end{align}
$(x:0\to\infty, u:1/2\to0)$. 
As $x\to\infty$, $e^{ikx} \sim u^{-ik/2}$.
The solutions are 
\begin{subequations}
\begin{align}
f_+ &=u^{-ik/2}(1-u)^{ik/2}{}_2F_{1}(\lam, 1-\lam,1-ik ,u)~,\\
f_- &=u^{ik/2}(1-u)^{-ik/2}{}_2F_{1}(\lam, 1-\lam,1+ik ,u)~.
%
\end{align}
\end{subequations}

The function $f_+$ is related to the Jost function (\appen{preliminaries}):
\begin{subequations}
\begin{align}
\Fp &= \lim_{x\to0} f_+ \\
&= 2^{ik}
\sqrt{\pi} \frac{ \Gamma(1-ik) }{ \Gamma\left(\frac{2-ik-\lam}{2}\right)\Gamma\left(\frac{1-ik+\lam}{2}\right)}~,
%
\end{align}
\end{subequations}
where we use 
\begin{align}
%
{}_2F_1(a,1-a,c;1/2) &= \frac{ 2^{1-c}\Gamma(c)\sqrt{\pi} }{ \Gamma(\frac{a+c}{2})\Gamma(\frac{c-a+1}{2}) }~.
%
\end{align}
The $S$-matrix is given by
\begin{align}
S = 2^{-2ik}
\frac{ \Gamma(\frac{2-ik-\lam}{2})\Gamma(\frac{1-ik+\lam}{2})\Gamma(1+ik) }{ \Gamma(\frac{2+ik-\lam}{2})\Gamma(\frac{1+ik+\lam}{2})\Gamma(1-ik) }~.
%
\end{align}
Note that 
\begin{subequations}
\begin{align}
\varphi &\propto \Fp f_- -\Fm f_+ \\
&=f_+(0)f_-(x) - f_-(0)f_+(x) 
%
\end{align}
\end{subequations}
so that $\varphi$ satisfies $\varphi(0)=0$. 

The $S$-matrix has poles at
\begin{subequations}
\begin{alignat}{2}
\text{Pole 1: } &\Gamma\left(\frac{2-ik-\lam}{2}\right) &&\to k=i(\lam-2-2\npa)~, \\
\text{Pole 2: } &\Gamma\left(\frac{1-ik+\lam}{2}\right) &&\to k = -i(\lam+1+2\npb)~, \\
\text{Pole 3: } &\Gamma(1+ik) &&\to k=i(\npc+1)~,
%
\end{alignat}
\end{subequations}
($\npa,\npb,\npc=0,1,\cdots$). The last one is the redundant poles in the upper-half plane. 

The $S$-matrix has zeros at
\begin{subequations}
\begin{alignat}{2}
\text{Zero 1: } & \Gamma\left(\frac{2+ik-\lam}{2}\right) &&\to k= -i(\lam-2-2\nza)~, \\
\text{Zero 2: } & \Gamma\left(\frac{1+ik+\lam}{2}\right) &&\to k = i(\lam+1+2\nzb)~, \\
\text{Zero 3: } & \Gamma(1-ik) &&\to k=-i(\nzc+1)~,
%
\end{alignat}
\end{subequations}
($\nza,\nzb,\nzc=0,1,\cdots$). The last one is the redundant zeros in the lower-half plane. 

For the details of the pole-skipping analysis, see \appen{details}. Again, we analyze the pole-skipping in the upper-half $k$-plane. Zero 3 is located in the lower-half plane, so we do not consider it. The potential remains the same under $\lam\to1-\lam$, so we consider $\lam\geq1/2$.

Also, summing the arguments of the Gamma functions for Pole 1 and Zero 2 give
\begin{align}
(2-ik-\lam)+(1+ik+\lam)=3~,
%
\end{align}
so they cannot be negative simultaneously, namely Pole 1 and Zero 2 never appear simultaneously. Similarly, Pole 2 and Zero 1 never appear simultaneously. The remaining combinations are
\begin{enumerate}
\item[(i)] Pole 1 and Zero 1: 
\begin{align}
k=0~, \quad\lam=2\npa+2~.
%
\end{align}
The pole-skipping occurs in physical region. 

\item[(ii)] Pole 2 and Zero 2: the pole-skip does not occur in the region $\lam\geq1/2$.

\item[(iii)] Pole 3 and Zero 1/Pole 3 and Zero 2: 
\begin{align}
k= ni~, \quad \lam= 1, \cdots, n~, 
%
\end{align}
($n=1,2\cdots$). The pole-skipping occurs at discrete values of imaginary $k$, and $k$ is evenly spaced like the previous example. 
However, Pole 3 is a redundant pole, so the physical interpretation may be subtle.

\end{enumerate}

The $S$-matrix has slope-dependence near pole-skipping points. For example, near $(k,\lam)=(i,1)$,
\begin{align}
S = \frac{\delta k+i\delta\lam}{\delta k}~.
%
\end{align}

The potential is regular at $x=0$, so it is appropriate to look at the $x=\infty$ expansion. The solution $f_-$ is expanded as
\begin{align}
f_- \sim u^{ik/2} \left\{1+ \frac{k(ik+1)+2i\lam(\lam-1)}{2(k-i)}u + \cdots \right\}~.
%
\end{align}
The $O(u)$ term takes the form 0/0 at the first pole-skipping point $(k, \lam)=(i,1)$.

\section{Holographic pole-skipping}\label{sec:holography}

According to holographic duality, a field theory (boundary theory) is equivalent to a classical gravitational theory (bulk theory). However, two theories live in different spacetime dimensions. The field theory typically lives in the 4-dimensional spacetime whereas the gravitational theory lives in the 5-dimensional spacetime. For a finite-temperature field theory, one considers the bulk spacetime with a black hole. A boundary theory operator corresponds to a bulk theory field. For example, a scalar operator $\calO$ on the boundary corresponds to a scalar field $\phi$ in the bulk. In order to obtain the finite-temperature Green's function for $\calO$, one solves the scalar field equation in the black hole background. 

Refs.~\cite{Grozdanov:2019uhi,Blake:2019otz,Natsuume:2019xcy} have shown that Green's functions are not unique at pole-skipping points (see also Refs.~\cite{Grozdanov:2017ajz,Blake:2018leo,Grozdanov:2018kkt,Natsuume:2019sfp}). Since then, various aspects of the pole-skipping have been investigated (see, \eg, Refs.~\cite{Natsuume:2019vcv,Wu:2019esr,Balm:2019dxk,Ceplak:2019ymw,Ahn:2019rnq,Ahn:2020bks,Abbasi:2020ykq,Jansen:2020hfd,Ramirez:2020qer,Ahn:2020baf,Natsuume:2020snz,Kim:2020url,Sil:2020jhr,Ceplak:2021efc,Jeong:2021zhz}).
This phenomenon can be seen in various Green's functions, \eg, the Green's functions for conserved quantities such as energy density, momentum density, and charge density. 

The pole-skipping in holographic duality and the pole-skipping in quantum mechanics are not mere analogy. A field equation in a black hole background can be written as an effective one-dimensional \schrodinger\ problem and the effect of the curved spacetime is encoded in an effective potential (see, \eg, \cite{BHP}). The pole-skipping in quantum mechanics is mathematically similar to the pole-skipping in holographic duality.

More explicitly, consider the following metric:
\begin{align}
ds^2 &= - F(r)dt^2+ \frac{dr^2}{F(r)} +\cdots~.
\end{align}
For simplicity, we set the horizon radius $r_0=1$. Near the horizon, 
\begin{align}
F(r)\sim 4\pi T(r-1)~.
\label{eq:near_horizon}
\end{align}
In the tortoise coordinate $r_*$,
\begin{subequations}
\begin{align}
ds^2 &=F(-dt^2+dr_*^2)+\cdots~,\\
dr_* &:= \frac{dr}{F}~.
%
\end{align}
\end{subequations}
For example, consider a minimally-coupled scalar field $\phi$:
\begin{align}
0=(\nabla^2-m^2)\phi~.
%
\end{align}
Consider the perturbation of the form $\phi(r)e^{-i\omega t+iqz}$. By redefining $\phi$ appropriately $\phi =:G(r)\tilphi(r)$, the field equation reduces to the \schrodinger\ form:
\begin{subequations}
\label{eq:effective}
\begin{align}
0 &= - \del_*^2 \tilphi +V \tilphi - \omega^2\tilphi~, 
 \\
V &=F(m^2+\cdots)~.
%
\end{align}
\end{subequations}
In order to obtain the retarded Green's function, one imposes the ``incoming-wave" boundary condition at the horizon:
\begin{align}
\tilphi \sim e^{-i\omega r_*}~, \quad (r_*\to-\infty)~.
%
\end{align}

However, the tortoise coordinate $r_*$ covers $-\infty<r_*<\infty$ (for asymptotically flat spacetimes) whereas our quantum mechanics problem is a half-line problem with $x>0$. In order to compare the black hole problem with the quantum mechanics problem, it is more appropriate to write \eq{effective} as a half-line problem. This is possible by a coordinate transformation%
\footnote{We transformed the field equation first in terms of the tortoise coordinate and transformed the equation again in terms of $x$. But this is unnecessary. One may introduce $x:=r-1$ and rewrite the field equation in the \schrodinger\ form. 
}:
\begin{align}
dr_*= \frac{dx}{F}~,
%
\end{align}
where $x$ and $r$ are related by $x:=r-1$. Near the horizon $x\to0$, $x\sim e^{4\pi T r_*}$. By redefining 
\begin{align}
\tilphi =: \sqrt{\frac{4\pi T}{F}} \psi~.
%
\end{align}
\eq{effective} is transformed as
\begin{subequations}
\begin{align}
0&=-\del_x^2\psi+U\psi~, \\
U &=\frac{V-\omega^2}{F^2} +\frac{1}{\sqrt{F}}\del_x^2 \sqrt{F} \\
&\sim \frac{\nu^2-1/4}{x^2}+O(x^{-1})~, 
\label{eq:effective_angular} \\
\nu &:= -\frac{i\omega}{4\pi T}~.
%
\end{align}
\end{subequations}
The incoming-wave boundary condition is transformed as our quantum mechanics boundary condition:
\begin{align}
\psi \sim x^{1/2+\nu}~, \quad (x\to0)~.
%
\end{align}
Thus, the black hole problem with $-i\omega$ reduces to the half-line quantum mechanics problem with the angular momentum $\nu=l+1/2$. In the black hole problem, the pole-skipping points are located at negative imaginary Matsubara frequencies $\omega=-(2\pi T)ni$, where $n=1,2,\cdots$.
As a quantum mechanics problem, this is translated into the pole-skipping at discrete negative angular momentum $\nu=-n/2$.
 In other words, \textit{the holographic pole-skipping has the universality because it reduces to a quantum mechanics problem with angular momentum.} 

\section{Discussion}\label{sec:discussion}

\begin{itemize}

\item 
The pole-skipping has the universality $\nu=-n/2$. The holographic pole-skipping also has the universality $\omega=-(2\pi T)ni$. This is because the holographic pole-skipping all reduces to quantum mechanics problems with angular momentum. Various other form of potentials is possible in quantum mechanics, and it would be interesting to study the other potentials, but probably there is no universality for generic potentials. 

\item
In this paper, we examine the potentials where analytic solutions are available, but our results suggest that an analytic solution is not really necessary to locate pole-skipping points. As long as one can construct power-series solutions, one can locate pole-skipping points, so one can analyze various other potentials as well. 

\item
We examine the nonuniqueness of the $S$-matrix and the nonuniqueness of the wave function. They are related to each other. The $S$-matrix is given by the Jost functions $\Fpm$. The functions $\Fpm$ can be written by the Wronskian of $\varphi$ and $f_\pm$ (\appen{preliminaries}). The former is the wave function which satisfies the $x=0$ boundary condition, and the latter is the wave function which satisfies the $x=\infty$ boundary condition. In this sense, the nonuniqueness of the $S$-matrix and the wave function is related, but they may not be equivalent.

Also, we use the power-series expansion to construct the wave functions. It is not entirely clear if the power-series expansion can locate all pole-skipping points. There may be some pole-skipping points which may not be found in the power-series expansion. ($k=0$ pole-skipping points may be such examples.) Conversely, some pole-skipping points found in the power-series expansion may not be the correct ones. The expansion is a locally obtained result whereas the wave function itself and the $S$-matrix are globally obtained results. It is desirable to find the necessary and sufficient condition for pole-skipping points. 

\item
So far, we consider idealistic potentials. For example, we consider the Coulomb potential which diverges at $x=0$, but a realistic potential does not really diverge. Similarly, we consider exponential type potentials which do not have a finite support. A realistic potential may have a finite support. In other words, we may truncate the potential in reality. The cutoff may affect the pole-skipping. We do not have conclusive answers, but some pole-skipping points seem to be sensitive to the deformations (\appen{cutoff}). 

However, the situation is different for the holographic pole-skipping or the perturbation problem in a black hole background. In this case, it is natural not to impose such a cutoff. The presence of the horizon requires that \eq{near_horizon} is valid at $x=0$. The effective potential $U$ then must behave as \eq{effective_angular} as $x\to0$. 

\item
It would be interesting to explore whether the pole-skipping has an observable consequence. 
In our examples, the pole-skipping in the $x=0$ expansion does not occur in physical region in the sense that 
$\nu<-1$. 
The pole-skipping in the $x=\infty$ expansion may occur in physical region. But it turns out that it corresponds to a redundant pole, so the physical interpretation is subtle. Namely, it may be sensitive to an IR cutoff. On the other hand, the pole-skipping may occur when $k=0$, and it does not seem to be problematic. Note that we are not completely excluding the pole-skipping in physical region. There may be some potentials which have the pole-skipping in physical region. It is interesting to find such potentials. 

\item 
As a quantum mechanics problem, the pole-skipping often occurs in unphysical region. However, as the corresponding black hole problem, the pole-skipping can have physical interpretations. In the holographic pole-skipping, $\omega$ is pure imaginary and the wave number $q$ is complex in general at pole-skipping points. So, the physical interpretation is not straightforward as well. However, for the energy-density Green's function, the $\omega=(2\pi T)i$ pole-skipping is related to the quantum many-body chaos, and the imaginary $q$ has the interpretation as the butterfly velocity \cite{Grozdanov:2017ajz,Blake:2018leo}. The quantum many-body chaos has been widely discussed in holography \cite{Shenker:2013pqa,Roberts:2014isa,Roberts:2014ifa,Shenker:2014cwa,Maldacena:2015waa}. Also, for the charge density Green's function, the pole-skipping occurs in physical region (lower-half $\omega$-plane and real $q$). In practice, one should be able to detect the pole-skipping by tuning the wave number $q$.

\item
Finally, the pole-skipping is not limited to holographic duality and quantum mechanics. The pole-skipping in a broad sense may be possible for the other areas such as electromagnetism.

\end{itemize}

\section*{Acknowledgments}


MN would like to thank Akinobu Dote, Osamu Morimatsu, and Izumi Tsutui for useful discussions. 
This research was supported in part by a Grant-in-Aid for Scientific Research (17K05427) from the Ministry of Education, Culture, Sports, Science and Technology, Japan. 


\appendix 

\section{Preliminaries}\label{sec:preliminaries}

First, it is convenient to introduce two sets of independent solutions:
\begin{enumerate}
\item The solutions which satisfy boundary conditions at $x=0$: suppose that the \schrodinger\ equation is approximated as
\begin{align}
0 \sim \del_x^2\psi - \frac{\nu^2-1/4}{x^2} \psi~, \quad (x\to0)~.
%
\end{align}
%
%
One can define two independent solutions :
\begin{align}
\varphi(\pm\nu) \to x^{1/2\pm\nu}~, \quad (x\to0)~.
\label{eq:bc_x=0}
\end{align}
The function $\varphi(\nu)$ is called the ``regular solution."
\item
The solutions which satisfy boundary conditions at $x=\infty$: if the potential decays fast enough asymptotically, one can define two independent solutions:
\begin{align}
f_\pm \to e^{\pm ikx}~, \quad (x\to\infty)~.
%
\end{align}
The functions $f_\pm$ are called the ``irregular solutions" or ``Jost solutions."
\end{enumerate}
The Wronskian $W[f_+,f_-]$ is independent of $x$, so it can be evaluated in the limit $x\to\infty$:
\begin{align}
W[f_+,f_-] = f_+f_-' - f_+'f_- = -2ik~.
\label{eq:w1}
\end{align}
Similarly, using the boundary condition \eqref{eq:bc_x=0}, one gets
\begin{align}
W[\varphi(\nu),\varphi(-\nu)] = -2\nu~.
\label{eq:w2}
\end{align}

The solutions $\varphi(\pm\nu)$ and the solutions $f_\pm$ are the solutions of the same \schrodinger\ equation, so
\begin{align}
\varphi(\nu) = af_- + bf_+~.
%
\end{align}
Then, 
\begin{subequations}
\begin{align}
&W[f_+,\varphi(\nu)] = aW[f_+,f_-] = -2ika~, \\
\to\, & 
a=\frac{1}{-2ik}W[f_+,\varphi(\nu)] =: \frac{1}{-2ik}\Fp(\nu)~.
%
\end{align}
\end{subequations}
Similarly, 
\begin{align}
b&=\frac{1}{2ik}W[f_-,\varphi(\nu)] =: \frac{1}{2ik}\Fm(\nu)~.
%
\end{align}
The functions $\mathcal{F}_\pm$ are called Jost functions.
Then,
\begin{subequations}
\label{eq:varphi}
\begin{align}
\varphi(\nu) &= -\frac{1}{2ik}\{\Fp(\nu) f_-(x) - \Fm(\nu) f_+(x) \} \\
&\sim -\frac{1}{2ik}\{\Fp(\nu) e^{-ikx} - \Fm(\nu) e^{ikx}\}~,
%
\end{align}
\end{subequations}
so the $S$-matrix is given by
\begin{align}
S = \frac{\Fm(\nu)}{\Fp(\nu)}~.
%
\end{align}
Likewise, 
\begin{align}
\varphi(-\nu) &= -\frac{1}{2ik}\{\Fp(-\nu) f_-(x) - \Fm(-\nu) f_+(x) \}~.
%
\end{align}

The relation between $\varphi(\pm\nu)$ and $f_\pm$ are written in a matrix form:
\begin{align}
\begin{pmatrix} 
  \varphi(\nu) \\ \varphi(-\nu)
\end{pmatrix}
  & =\frac{1}{2ik} \begin{pmatrix} 
    \Fm(\nu) & -\Fp(\nu)  \\
    \Fm(-\nu) & -\Fp(-\nu) 
  \end{pmatrix}
\begin{pmatrix} 
    f_+ \\ f_-
\end{pmatrix}~.
%
\end{align}
Its inverse is given by
\begin{align}
\begin{pmatrix} 
    f_+ \\ f_-
\end{pmatrix}
  & =\frac{1}{2\nu} \begin{pmatrix} 
    -\Fp(-\nu) & \Fp(\nu)  \\
    -\Fm(-\nu) & -\Fm(\nu) 
  \end{pmatrix}
\begin{pmatrix} 
  \varphi(\nu) \\ \varphi(-\nu)
\end{pmatrix}~.
\label{eq:f_by_varphi}
\end{align}
Here, we use the relation
\begin{align}
\Fp(\nu)\Fm(-\nu) - \Fp(-\nu)\Fm(\nu) =4ik\nu
%
\end{align}
using Eqs.~\eqref{eq:w1} and \eqref{eq:w2}.

One can obtain $\mathcal{F}_\pm$ from \eq{varphi}. Alternatively, one can obtain them from $f_\pm$. Using \eq{f_by_varphi}, one obtains
\begin{subequations}
\begin{align}
2\nu f_+ &= \Fp(\nu)\varphi(-\nu) - \Fp(-\nu)\varphi(\nu) 
\\
&\sim \Fp(\nu)x^{1/2-\nu}~, \quad (x\to0, \text{Re}~\nu>0) 
\\
\to \Fp(\nu) &= 2\nu \lim_{x\to0} x^{\nu-1/2} f_+~.
%
\end{align}
\end{subequations}

\section{Details of pole-skipping analysis}\label{sec:details}

The pole-skipping analysis is straightforward, but some care is necessary. In some cases, an additional pole (or zero) appears from the other Gamma functions. In such a case, the $S$-matrix has poles (or zeros) of degree 2, and the nonuniqueness disappears. One has to exclude such cases.

\subsection{$1/\sinh^2x$}

The poles and zeros are given by
\begin{subequations}
\begin{alignat}{3}
\text{Pole 1: } & k = -i \left(\npa+\nu+\frac{1}{2}\right)~,  
\\
\text{Pole 2: } & k = i(\npb+1)~,
\\
%
\text{Zero 1: } & k = i \left(\nza+\nu+\frac{1}{2}\right)~,
 \\
\text{Zero 2: } & k = -i(\nzb+1)~,
%
\end{alignat}
\end{subequations}
It is enough to analyze the pole-skipping in the upper-half $k$-plane. Zero 2 is located in the lower-half $k$-plane, so we do not consider it. 

\begin{enumerate}
\item[(i)] Pole 1 and Zero 1:
The pole-skipping points are given by
\begin{subequations}
\label{eq:sh_p1z1}
\begin{align}
\nu &=-\frac{\npa+\nza+1}{2}~, \\
k &=\frac{i}{2}(\nza-\npa)~.
\end{align}
\end{subequations}
However, when $k=im~ (m=\mathbb{Z}^+)$, it coincides with Pole 2. To exclude it, set
\begin{align}
\npa+\nza=2n+1~, \quad (n=0,1,\cdots)~.
%
\end{align}
Then, 
\begin{subequations}
\begin{align}
\nu &= -n-1~, \\
k &= i\left(n-\npa+\frac{1}{2}\right)~, 
%
\end{align}
\end{subequations}
($\npa <n+1/2$).
In addition, as a special case $\npa=\nza$ of \eq{sh_p1z1},
\begin{align}
k=0~, \quad \nu=-\npa-\frac{1}{2}~.
%
\end{align}

\item[(ii)] Pole 2 and Zero 1:
The pole-skipping points are given by
\begin{subequations}
\begin{align}
\nu &= \npb-\nza+\frac{1}{2} = n-\frac{1}{2}~, \\
k &=i(\npb+1)~,
%
\end{align}
\end{subequations}
where $ n:=\npb-\nza+1\leq \npb+1$.
Pole 1 should not appear simultaneously, so $\nu+1/2-ik=n+\npb+1>0$, or $-\npb-1<n$. 
Thus, $-\npb \leq n \leq \npb+1$. 

\end{enumerate}

\subsection{$1/\cosh^2x$}

The poles and zeros are given by
\begin{subequations}
\begin{alignat}{2}
\text{Pole 1: } & k=i(\lam-2-2\npa)~, \\
\text{Pole 2: } & k = -i(\lam+1+2\npb)~, \\
\text{Pole 3: } & k=i(\npc+1)~,\\
\text{Zero 1: } & k= -i(\lam-2-2\nza)~, \\
\text{Zero 2: } & k = i(\lam+1+2\nzb)~, \\
\text{Zero 3: } & k=-i(\nzc+1)~,
%
\end{alignat}
\end{subequations}
Zero 3 is located in the lower-half $k$-plane, so we do not consider it. 
The potential remains the same under $\lam\to1-\lam$, so we consider $\lam\geq1/2$. 
Pole 1 and Zero 2 never appear simultaneously. Similarly, Pole 2 and Zero 1 never appear simultaneously. The remaining combinations are
\begin{enumerate}
\item[(i)] Pole 1 and Zero 1: 
\begin{subequations}
\label{eq:ch_p1z1}
\begin{align}
k &= i(\nza-\npa)~, \\
\lam &= \npa+\nza+2~.
%
\end{align}
\end{subequations}
Pole 2 does not appear simultaneously with Zero 1, and Zero 2 does not appear simultaneously with Pole 1, but Pole 3 ($ik=-\npc-1$) appears simultaneously (if $\npa\neq\nza$), so the pole-skipping does not occur. 
This leaves the case $\npa=\nza$ only:
\begin{align}
k=0~, \quad\lam=2\npa+2~.
%
\end{align}

\item[(ii)] Pole 2 and Zero 2: 
\begin{subequations}
\label{eq:ch_p2z2}
\begin{align}
k &= i(\nzb-\npb)~, \\
\lam &= -\npb-\nzb-1~,
%
\end{align}
\end{subequations}
but $\lam<0$, so we do not consider this case further.
%
%

\item[(iii)] Pole 3 and Zero 1: 
\begin{subequations}
\begin{align}
k &= i(\npc+1)~, \\
\lam &= 2\nza-\npc+1~.
%
\end{align}
\end{subequations}
Pole 2 does not appear simultaneously with Zero 1. For Zero 2, $1+ik+\lam=2\nza-2\npc+1$ which is an odd integer. The Gamma function argument $(1+ik+\lam)/2$ is not an integer, so Zero 2 does not appear simultaneously. But Pole 1 should not appear simultaneously, so $2-ik-\lam=2\npc-2\nza+2>0$, or $0\leq \nza \leq\npc$ (we impose $\kappa\geq1/2$ later). 

For example,
\begin{subequations}
\begin{align}
k=i~, \quad
&\lam=1~, \\
k=2i~, \quad
& \lam = (0)~, 2~, \\
\cdots \quad & \cdots 
\nonumber
%
\end{align}
\end{subequations}

\item[(iv)] Pole 3 and Zero 2:
\begin{subequations}
\begin{align}
k &= i(\npc+1) \\
\lam &= -2\nzb+\npc
%
\end{align}
\end{subequations}
Pole 1 does not appear simultaneously with Zero 2. For Zero 1, $2+ik-\lam=2\nzb-2\npc+1$ which is an odd integer. The Gamma function argument $(2+ik-\lam)/2$ is not an integer, so Zero 1 does not appear simultaneously. But Pole 2 should not appear simultaneously, so $1-ik+\lam=2\npc-2\nzb+2>0$, or $0\leq\nzb\leq\npc$ (we impose $\kappa\geq1/2$ later).

For example,
\begin{subequations}
\begin{align}
k=i~, \quad
&\lam=(0)~, \\
k=2i~, \quad
& \lam = (-1)~, 1~, \\
\cdots \quad & \cdots 
\nonumber
%
\end{align}
\end{subequations}

\end{enumerate}
Pole 3-Zero 1 and Pole 3-Zero 2 are summarized as
\begin{align}
k= ni~, \quad \lam= 1, \cdots, n~, \quad (n=1,2\cdots)~,
%
\end{align}
where we restrict to $\lam\geq1/2$. 

\section{Cutoff dependence}\label{sec:cutoff}

In this Appendix, we truncate the potentials at large $x$ and small $x$ and see the effect on the pole-skipping analysis. We do not have conclusive answers, but some pole-skipping points seem to be sensitive to the deformations. 


\subsection{IR cutoff}

For an exponential potential, we impose an ``IR cutoff," namely we truncate the potential at large $x$: 
\begin{align}
V_r(x) = \left\{ 
\begin{array}{ll}
V(x)~, & x<R~,\\
0~, & x>R~.
\end{array}
\right.
%
\end{align}
A characteristic feature of an exponential potential is the existence of redundant poles. At a redundant pole, $\Fm=\infty$, so the asymptotic expansion \eqref{eq:asymptotic} is not really valid. If we impose the cutoff, the asymptotic expansion should be now valid, so there should be no redundant poles. On the other hand, the cutoff should not affect the bound states. Cutting off the potential far away should not change the wave functions of bound states which are localized at small $x$. One can show this explicitly in simple examples (see, \eg, Refs.~\cite{redundant}).

The IR cutoff affects our pole-skipping analysis. For the $1/\sinh^2 x$ potential, Pole 2 and Zero 2 are redundant poles and zeros, respectively. If one imposes an IR cutoff, they should disappear, and the corresponding pole-skipping points also should disappear. We are not sure if the IR cutoff affects the remaining pole-skipping points from Pole 1-Zero 1. 

\subsection{UV cutoff}

For the Coulomb potential and the $1/\sinh^2 x$ potential, we impose an ``UV cutoff" at small $x$: 
\begin{align}
V_r(x) = \left\{ 
\begin{array}{ll}
V(x)~, & x>a~,\\
V(a)=V_0~, & x<a~.
\end{array}
\right.
%
\end{align}
Below we use formal properties of Jost functions (see \appen{preliminaries}).

The renormalized Jost solution $f_+^r(x)$ is given by
\begin{align}
f_+^r = \left\{ 
\begin{array}{ll}
f_+~, & x>a~,\\
C_- e^{-ik_0x} + C_+ e^{ik_0x}~, & x<a~,
\end{array}
\right.
%
\end{align}
where $k_0^2=k^2-V_0$. As usual, we impose the conditions that $f_+^r$ and its derivative are continuous at $x=a$. 

The potential $V_r(x)$ is regular at the origin. 
The renormalized solution $\varphi^r$ near the origin is then given by
\begin{align}
\varphi^r(x)= \frac{1}{k_0}\sin{k_0x}~,
%
\end{align}
where we impose the boundary condition $\varphi(0)=0, \varphi'(0)=1$ which corresponds to the case $\nu=1/2$ in \eq{bc_x=0}.

The renormalized Jost function $\Fp^r(\nu)$ is 
\begin{align}
\Fp^r(\nu) = W[f_+^r(x),\varphi^r(x)]~.
%
\end{align}
The Wronskian is independent of $x$, so one can evaluate it at $x=a$. Moreover, the function $f_+^r$ and its derivative is continuous at $x=a$, so one can replace $f_+^r$ by $f_+$:   
\begin{align}
\Fp^r(\nu) = W[f_+(a),\varphi^r(a)]~.
%
\end{align}
The function $f_+$ is rewritten in terms of $\varphi(x,\pm\nu)$ as \eq{f_by_varphi}:
\begin{align}
f_+ &= \frac{1}{2\nu}\{-\Fp(-\nu) \varphi(x,\nu) + \Fp(\nu) \varphi(x,-\nu) \}~.
%
\end{align}
Thus, 
\begin{align}
\Fp^r(\nu) =&  \frac{1}{2\nu} \{ -\Fp(-\nu) W[\varphi(a,\nu),\varphi^r(a)] 
\nonumber \\
&+ \Fp(\nu) W[\varphi(a,-\nu),\varphi^r(a)] \}~.
%
\end{align}
The functions $\varphi(a,\nu)\sim a^{1/2+\nu}$ and $\varphi(a,-\nu)\sim a^{1/2-\nu}$. When $\text{Re}~\nu>0$, the second term dominates, so $\Fp^r(\nu) \propto \Fp(\nu)$. Namely, the renormalized Jost function is given by the original Jost function. When $\text{Re}~\nu<0$, the first term dominates, so $\Fp^r(\nu) \propto \Fp(-\nu)$. Then, the renormalized Jost function is given by the original Jost function with a positive value of $\nu$. In both cases, the renormalized Jost function is determined by the original Jost function with $\text{Re}~\nu>0$.

Consequently, the $S$-matrix shows the same behavior. Namely, the renormalized $S$-matrix $S^r$ is determined by the $\text{Re}~\nu>0$ region of the original $S$-matrix. In our examples, the pole-skipping points are located at $\nu<0$, so these pole-skipping points disappear in the presence of the UV cutoff.

\end{document}